\pdfoutput=1
\documentclass[fleqn,usenatbib]{mnras}

\usepackage{newtxtext,newtxmath}
\usepackage[T1]{fontenc}

\DeclareRobustCommand{\VAN}[3]{#2}
\let\VANthebibliography\thebibliography
\def\thebibliography{\DeclareRobustCommand{\VAN}[3]{##3}\VANthebibliography}

\usepackage{graphicx}	% Including figure files
\usepackage{amsmath}	% Advanced maths commands
\usepackage[dvipsnames]{xcolor}

\newcommand{\ppxf}{\textsc{pPXF}}
\newcommand{\aFe}{[$\alpha/\mathrm{Fe}$]}
\newcommand{\kms}{\mathrm{km\,s}^{-1}}

\title[Kinematically isolating ICL]{Kinematically isolating intracluster light in AS1063: a distinct stellar population tracing the dark matter core}

\author[C. Bellhouse et al.]{
Callum Bellhouse,$^{1}$\thanks{E-mail: icl-group@nottingham.ac.uk}
Steven Bamford,$^{1}$
Nina Hatch,$^{1}$
Tutku Kolcu,$^{1}$
Johan Richard$^{2}$
\\
$^{1}$School of Physics and Astronomy, University of Nottingham, University Park, Nottingham NG7 2RD, UK\\
$^{2}$Univ Lyon, Univ Lyon1, ENS de Lyon, CNRS, Centre de Recherche Astrophysique de Lyon UMR5574, 69230 Saint-Genis-Laval, France\\
}

\date{Submitted to MNRAS 4 September 2026}

\pubyear{\the\year{}}

\begin{document}
\label{firstpage}
\pagerange{\pageref{firstpage}--\pageref{lastpage}}
\maketitle

\begin{abstract}
We present a technique using integral-field spectroscopy to kinematically isolate intracluster light (ICL) from the brightest cluster galaxy (BCG), enabling insights into their contrasting stellar populations and the inner dark matter halo profile. Applying this technique to MUSE data of Abell~S1063, we find that both components are old (${\sim}9$\,Gyr), but the ICL is more metal-poor than the BCG, implying distinct progenitor pathways. Within 100\,kpc, the ICL comprises 55\,per cent of the BCG+ICL stellar mass -- substantially more than inferred from photometric decompositions. The BCG and ICL possess distinct velocity distributions, which remain nearly constant with radius, with an ICL velocity dispersion more than twice that of the BCG, and a line-of-sight velocity offset of $86\pm18\,\kms$. Two-dimensional S\'ersic fits yield an ICL centroid offset of $5.2\pm1.5$\,kpc from the BCG. These phase-space offsets are indicative of the BCG oscillating in a broad potential minimum. Furthermore, the ICL surface-brightness profile matches that of strong-lensing estimates assuming a cored halo. Simulations indicate that the inner halo profile varies systematically for different dark matter models. Applied to larger samples, our method therefore offers a complementary route to testing the nature of dark matter, while simultaneously revealing the stellar origins of the ICL.
\end{abstract}


%%%%%%%%%%%%%%%%%%%%%%%%%%%%%%%%%%%%%%%%%%%%%%%%%%

%%%%%%%%%%%%%%%%% BODY OF PAPER %%%%%%%%%%%%%%%%%%

\section{Introduction}\label{sec:intro}

Intracluster light (ICL) is a diffuse stellar component of galaxy clusters, composed of stars that have been stripped from their host galaxies through tidal interactions and mergers, and are instead bound to the gravitational potential of the cluster as a whole \citep{Zwicky1951}. As a record of disrupted galaxies, the ICL provides insights into the hierarchical assembly of clusters \citep{Contini2014, Mihos2017}, encodes the timing and nature of past dynamical interactions \citep{Montes2018, DeMaio2020}, and closely traces the underlying dark matter distribution on large scales, potentially serving as a luminous proxy for dark matter in optical observations \citep{Montes2019, AlonsoAsensio2020}.

The central regions of clusters -- where the ICL spatially overlaps with the brightest cluster galaxy (BCG) -- are particularly important for understanding both cluster and BCG assembly. This region captures the buildup of both components through mergers and accretion \citep{Murante2007, Burke2015}, and provides a window into their stellar population differences, such as age and metallicity gradients \citep{Montes2014, Edwards2020, Barbosa2021}. Furthermore, detailed mapping of this transition region enables tests of dark matter models, as competing theories predict subtle differences in inner density profiles and dynamics \citep{Newman2013, Schaller2015, Robertson2019, Andrade2022}. Constraining the structure of the BCG--ICL interface is therefore essential for probing both baryonic and dark matter physics in cluster cores.

While the ICL and BCG can appear blended photometrically due to their continuous luminosity profiles, simulations show they occupy distinct dynamical states: ICL stars exhibit higher velocity dispersions and trace the cluster potential, whereas BCG stars are more centrally bound \citep{Dolag2010, Remus2017, Marini2021}. \citet{Pearce2026} explains the dichotomy as resulting from a strongly mass-dependent combination of dynamical friction, tidal stripping and violent relaxation. Observational evidence supports this separation through spectroscopy of integrated stellar light \citep{Kelson2002, Bender2015, Gu2020} and discrete tracers such as planetary nebulae \citep{Longobardi2015, Hartke2022}, both of which reveal rising velocity dispersion profiles consistent with an extended ICL component. Kinematics thus provide a physically motivated framework for separating the BCG and ICL, especially in the central regions.

Photometric methods remain widely used for ICL identification due to their simplicity and applicability to imaging surveys. These include surface-brightness thresholds \citep[e.g.][]{Feldmeier2004, Burke2015}, multi-component profile fitting \citep{Gonzalez2005, Kluge2021}, and non-parametric techniques such as wavelet decomposition, which isolate large-scale diffuse emission \citep{Ellien2019}. However, comparisons with simulations -- where the BCG and ICL can be defined kinematically -- show that photometric methods often underestimate the ICL fraction and are sensitive to model assumptions, projection effects, and data depth \citep{Rudick2011, Pillepich2018, Canas2020}. A recent study by \citet{Brough2024} applied a range of photometric techniques to mock observations from hydrodynamical simulations and found systematic biases across all methods. Multi-galaxy fitting yielded the most accurate results, while surface-brightness cuts and profile decompositions showed larger discrepancies. Empirical corrections were proposed to calibrate photometric ICL estimates to kinematic definitions. While some overlap exists between photometric and kinematic components in observed systems, they are not interchangeable, underscoring the limitations of relying solely on imaging data \citep{Longobardi2015, Barbosa2021, Hartke2022}.

A complementary -- and potentially transformative -- motivation for isolating the ICL is its use as a dynamical tracer of the inner halo, which could help to elucidate the nature of dark matter. While cold dark matter (CDM) is typically assumed by current simulations of galaxy formation, several alternative models -- including warm, mixed, fuzzy, and self-interacting dark matter (SIDM) -- can also match existing observations. A promising way to distinguish among these scenarios is to examine the inner regions of dark matter halos, since the additional physical processes in many alternative models alter the central structure of halos, producing `cored' mass distributions, with shallow central density profiles compared to the sharply-peaked, `cuspy', profiles formed in standard CDM \citep{Harvey2019}. Strong lensing is the current gold standard for measuring inner mass profiles \citep{Gassis2025}, but only in projection, and it is limited by our ability to model the detailed sub-structure of the mass distribution \citep{Roche2024, Perera2025}. X-rays are also useful, but they trace hot gas, which is complex to model and easily disturbed \citep{Sanders2020, XRISM2025}.

The gravitational potential within a galaxy cluster is set primarily by dark matter, so stars on higher-energy orbits that populate the extended ICL can, in principle, map that potential. Gravitational lensing already indicates that, in cluster outskirts, the isophotes of the diffuse stellar light closely follow the total mass distribution \citep{Montes2019}. Spectroscopically isolating the ICL throughout the cluster core would enable constraints on the inner density profile and, uniquely, simultaneous measurements of BCG--halo offsets in both projected position and line-of-sight velocity. Since motion of the BCG is less constrained within a cored halo, the joint distribution of such phase-space offsets across a sample of clusters is a predicted discriminator among dark matter models \citep{Harvey2019, Harvey2025}, provided the BCG and ICL can be separated robustly. To date, that separation has remained the principal obstacle to using ICL as a probe of halo cores.

In this work, we present a kinematic decomposition of a BCG and its associated ICL using integral field spectroscopy from the Multi Unit Spectroscopic Explorer (MUSE). Our target is Abell~S1063 (also known as RXC~J2248.7$-$4431 or MACS~2248.7$-$4431, hereafter AS1063), a massive galaxy cluster at $z = 0.3457$, with a velocity dispersion of $\sigma = 1380^{+26}_{-32}~\kms$ \citep{Mercurio2021} and mass $M_{200}\simeq 2.5\times10^{15}\,\mathrm{M}_\odot$ \citep{Rodriguez2024}. AS1063 is among the hottest known clusters in X-rays \citep[$kT > 11.5$~keV;][]{Gomez2012} and is one of the strongest Sunyaev--Zel'dovich detections in the South Pole Telescope survey \citep{Williamson2011}. Although AS1063 is classified as the most relaxed cluster in the Hubble Frontier Fields (HFF) sample \citep{Lotz2017}, it hosts significant substructure, including an X-ray offset from the galaxy distribution \citep{Gomez2012} and mass substructure identified in weak-lensing maps \citep{Gruen2013}. A large number of strongly lensed background sources allow the dark matter distribution to be estimated independently \citep[e.g.][]{Caminha2016, Beauchesne2025}. \citet{Montes2019} used deep \textit{HST} imaging to measure the ICL in AS1063 and found its morphology closely traces the cluster's mass distribution.

Leveraging the spatial resolution and depth of MUSE, we perform a full kinematic decomposition of the central stellar light. We show that BCG and ICL can be separated using their distinct stellar populations and kinematics, revealing both the stellar origins of the ICL and a phase-space offset relative to the BCG that -- together with the ICL surface-brightness profile -- provides insights regarding the inner dark matter halo.

This paper is structured as follows. In Sect.\,\ref{sec:data} we introduce the data and processing steps. Sect.\,\ref{sec:spectfit} describes the spectral fitting, including single- and two-component models. In Sect.\,\ref{sec:SBmaps} we present the decomposed surface brightness profiles, two-dimensional S\'ersic fits, and ICL fraction. In Sect.\,\ref{sec:discussion} we discuss the implications for our understanding of ICL and for constraining the properties of dark matter. Finally, in Sect.\,\ref{sec:conclusions} we summarise our conclusions. Throughout this paper we assume a standard $\Lambda$CDM flat cosmology with $\Omega_\mathrm{m}=0.3$ and $H_0=70\,\mathrm{km}\,\mathrm{s}^{-1}\,\mathrm{Mpc}^{-1}$.

\begin{figure*}
    \centering
    \includegraphics[width=\linewidth,trim=0 1cm 0 0,clip]{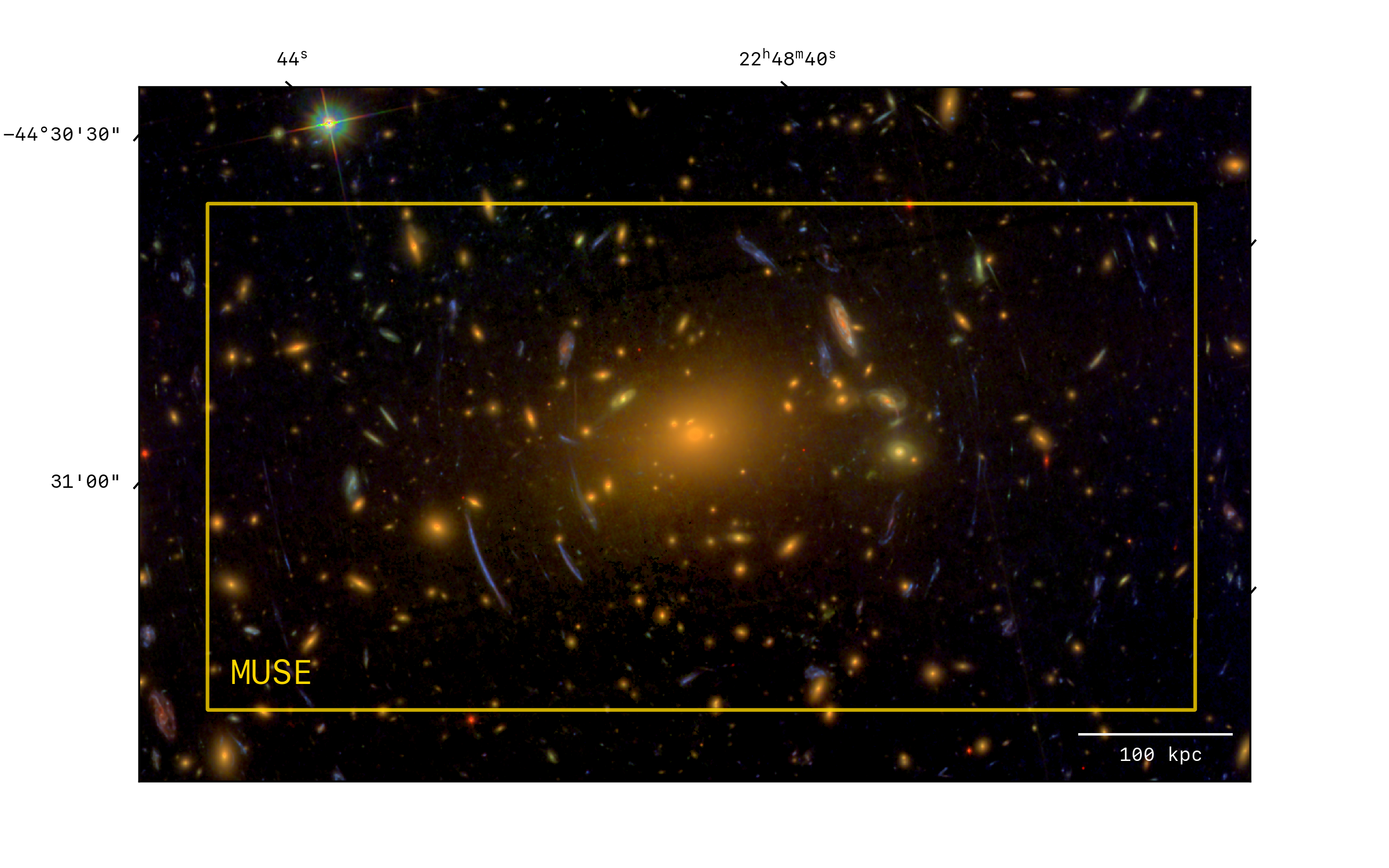}
    \caption{RGB image of AS1063 produced from \textit{HST} imaging in the F814W, F606W and F435W filters. The yellow box indicates the footprint of the combined MUSE pointings.}
    \label{fig:cluster_overview}
\end{figure*}

\begin{figure}
    \centering
    \includegraphics[width=\linewidth]{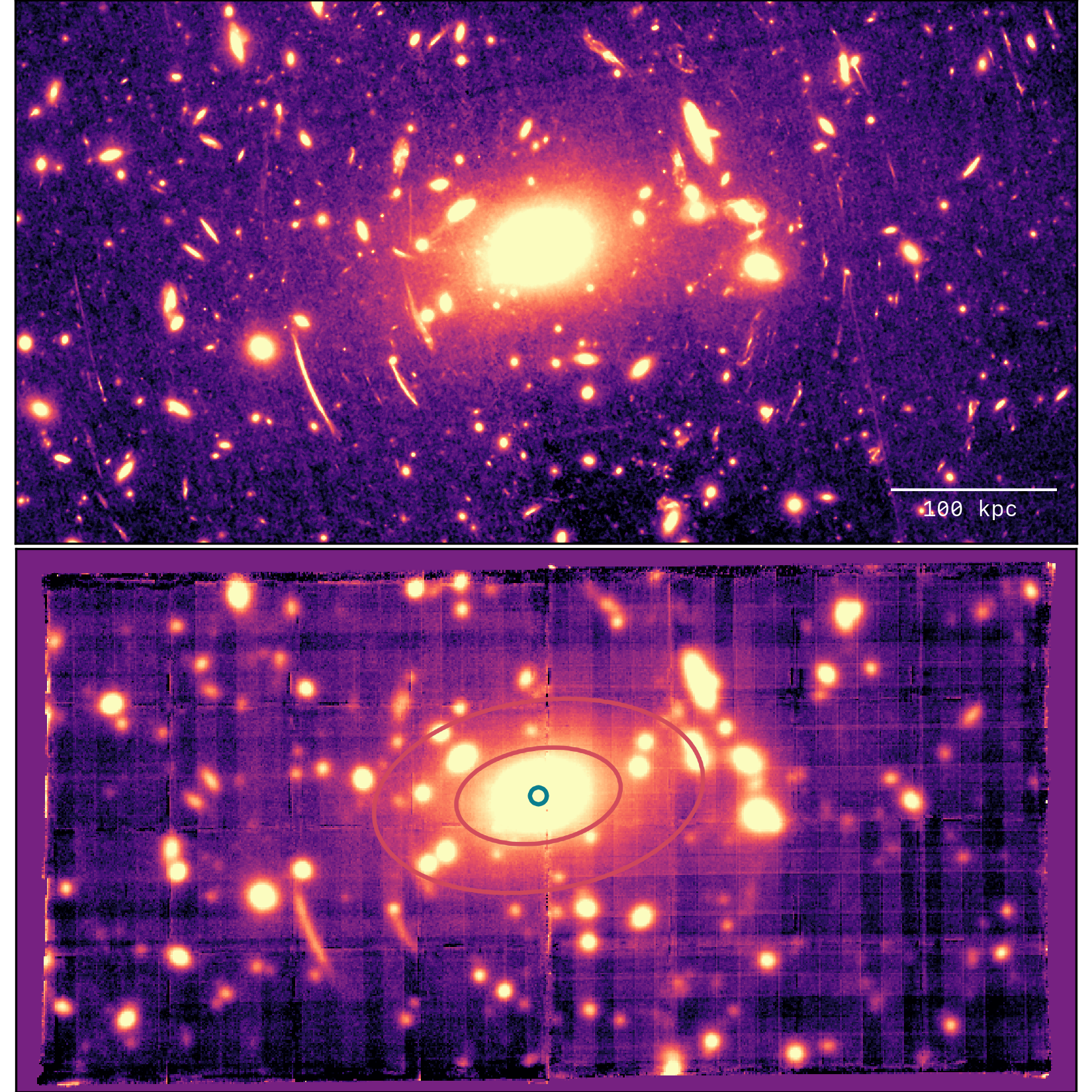}
    \caption{\textit{Top}: \textit{HST} F606W image, scaled to highlight the diffuse ICL emission.
    \textit{Bottom}: MUSE image produced by convolving the spectra with the response curve of the F606W filter. The red ellipses and blue circle show the bounds of the regions used to calculate the template ICL (50--100\,kpc elliptical annulus) and BCG (0--5\,kpc circular aperture) spectra respectively.}
    \label{fig:HST_MUSE_bins}
\end{figure}

\section{Data and processing}\label{sec:data}

We use integral-field spectroscopy from the MUSE instrument \citep{Bacon2010} on the Very Large Telescope (VLT). The observations comprise two pointings on AS1063 that cover the North-East and South-West regions \citep{Karman2015,Karman2017}, shown in Fig.\,\ref{fig:cluster_overview} and Fig.\,\ref{fig:HST_MUSE_bins}. Total on-source integration times are 4.8\,h (NE) and 3.1\,h (SW). For this analysis we reprocessed the observations with the standard pipeline \citep{Weilbacher2020}, but omitting the autocalibration step to preserve low-surface-brightness structure in the ICL.

We complement the spectroscopy with \textit{HST} Frontier Fields WFC3 imaging in F435W, F606W, and F814W, processed as in \citet{Lotz2017}. These images provide higher spatial resolution and deeper limiting surface brightness than MUSE and are used to build the source mask applied to the MUSE cube.

\subsection{Masking galaxies}\label{sec:masking}

To ensure we are measuring genuine ICL, it is important to exclude light from cluster satellites and line-of-sight sources. We therefore construct a mask from the deep \textit{HST} images in multiple stages. To start, we smooth the images with a Gaussian of $\sigma=0.55$\,arcsec to mimic the atmospheric smoothing (seeing) present in the MUSE data cube (this difference in spatial resolution can be seen when comparing panels of Fig.\,\ref{fig:HST_MUSE_bins}).
We subtract a two-dimensional median background from each \textit{HST} image, then detect and deblend sources based on a $5\sigma$ excess relative to the background-subtracted image. The resulting sources, except the BCG, form the first mask.
We next apply a stricter background subtraction (smaller box size) to capture spatial variations within the ICL region. Sources close to the BCG are again detected and deblended using a $5\sigma$ threshold. Additional source pixels identified in this pass (excluding the BCG) are added to the first mask.
We then repeat this entire procedure with the original unsmoothed \textit{HST} images, to locate smaller, fainter objects. We dilate this second mask by 0.72\,arcsec to account for the seeing in the MUSE data. The final mask is formed by combining all the masks, then reprojected to the MUSE pixel scale and orientation, and applied to the data cube to remove all sources of light except the BCG and diffuse light. The final masked \textit{HST} F606W image can be seen in Fig.\,\ref{fig:models}A.

\subsection{Background subtraction}\label{sec:background}

An initial background subtraction of the MUSE data cube was performed by the standard pipeline \citep{Weilbacher2020}, which successfully removed contamination from sky lines. To investigate and correct for any residual background, we take the following steps. We first apply the galaxy mask, then additionally mask a central region defined by an ellipse ($\epsilon=0.43$, $\mathrm{PA}=50$\,deg) centred on the approximate BCG position at $\mathrm{RA}=342.183$, $\mathrm{Dec}=-44.531$ with a semi-major axis of 200\,kpc. This ellipse was measured from the smoothed \textit{HST} F606W image using the isophotal fitting tool \texttt{AutoProf} \citep{Stone2021}. The shape of this extra mask is designed to conservatively exclude the region in which ICL is detectable. From the remaining unmasked regions, we create a white-light image by integrating the flux over all wavelengths of the spectra, then select the faintest 20\,per cent of spaxels in these regions. The spectra from these spaxels are averaged to construct a background spectrum. To verify the spatial uniformity of this background spectrum, we produce independent spectra from the four corners of the masked field and confirm no significant differences between them. The global background spectrum is fit with a 4th-order polynomial, which is then subtracted from every spectrum in the cube.

\subsection{Spatial voronoi binning}\label{sec:binning}

Because the ICL is extremely faint in individual spectra, we increase the signal-to-noise ratio (S/N) by combining spectra into spatial bins constructed using Voronoi tessellation, following \citet{Cappellari2003}. The two-pointing mosaic exhibits elevated noise along the central vertical join, which yields overly large bins if we target uniform S/N. Instead, we create Voronoi bins in which the square root of the total signal (flux integrated over all wavelengths) is constant. This yields a smoother distribution of bin sizes across the field of view, preserving higher spatial resolution in bright central regions, with larger bins at low surface brightness. The target flux is set by balancing adequate S/N in the faintest regions against retaining enough spatial bins to resolve azimuthal structure in the ICL.

\begin{figure}
    \centering
    \includegraphics[width=\linewidth]{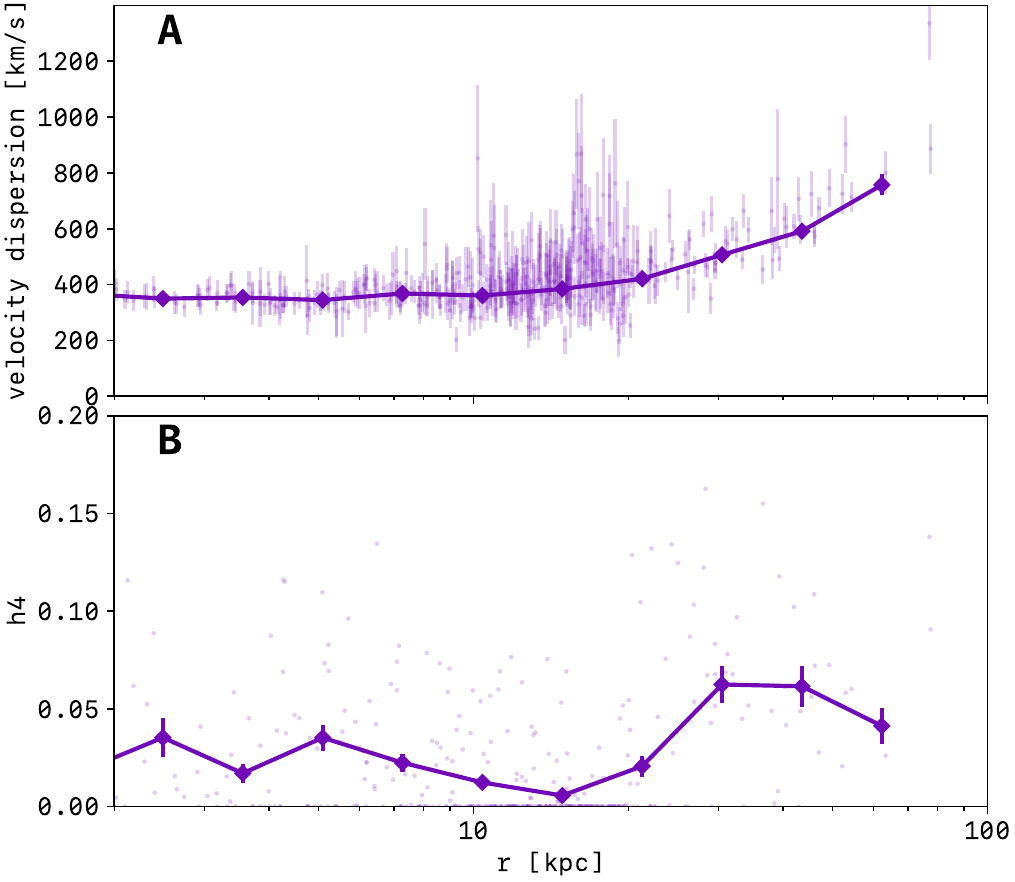}
    \caption{Velocity dispersion and $h_4$ moment profiles for \ppxf{} fits with a single kinematic component. Small points show the velocity dispersion (\textbf{A}) and the $h_4$ coefficient, or kurtosis, (\textbf{B}) of the LOSVD from the fit to the integrated spectrum in each spatial bin, versus the average radius of the spaxels within that bin. In panel A, error bars indicate the uncertainty reported by \ppxf{} for each bin. To ensure reliable kinematics, we only consider bins where the total (BCG+ICL) surface brightness is brighter than 25\,mag\,arcsec$^{-2}$. The large diamonds and lines show the error-weighted mean value in logarithmic bins of radius, with error bars indicating the standard error within each radial bin.}
    \label{fig:disp_h4}
\end{figure}

\section{Spectral fitting}\label{sec:spectfit}

\begin{figure*}
    \centering
    \includegraphics[width=0.75\hsize]{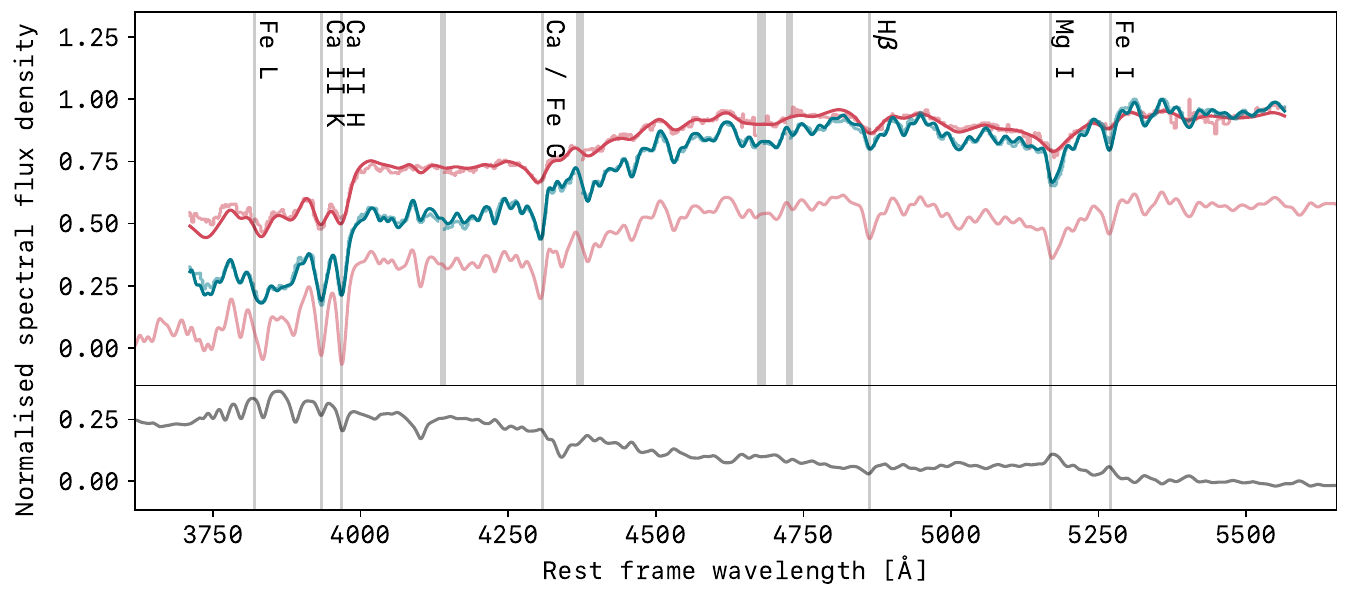}
    \caption{\textit{Top}: Integrated spectra of the BCG (blue) and ICL (red) within the regions shown in Fig.\,\ref{fig:HST_MUSE_bins}. The stepped lines show the MUSE spectra, with $2\times$ spectral binning. The solid curves show the \ppxf{} fits to the spectra, which represent the best-fit combination of stellar templates convolved with the LOSVD of the integrated spectra. The faint red line shows the best-fit template spectrum of the ICL convolved with the LOSVD of the BCG, offset down by 0.5 in normalized flux, to facilitate comparison between the components. The grey lines mark the wavelengths of prominent absorption lines, while grey shaded regions mark the locations of masked sky lines.
    \textit{Bottom}: The grey line shows the residual between the ICL and BCG best-fit templates, convolved with the LOSVD of the BCG.}
    \label{fig:spectra}
\end{figure*}

We employ \ppxf{} \citep{Cappellari2023} to fit each spectrum with, potentially, multiple components. Each component is formed from a linear combination of template spectra, convolved by a line-of-sight velocity distribution (LOSVD) with a Gauss--Hermite parametrization \citep{Cappellari2017}. One possible concern when modelling the ICL and BCG spectra is a potential mismatch between their $\alpha$-element abundances (\aFe) and those assumed by the stellar population templates. BCGs, for example, can display elevated $\alpha$-element abundances \citep{Gu2018}. Such mismatches may introduce systematic errors in metallicities and stellar ages derived from SSP fitting. While this is less problematic for an analysis focussed on kinematics, an accurate representation of the BCG and ICL spectra is important for decomposing the two components. We therefore modified \ppxf{} to use the SMILES SSPs \citep{Knowles2023} and simultaneously fit metallicity, stellar age, and \aFe{}.

\subsection{Single-component fits}\label{sec:singlecomp}

We first fit the spectrum from each spatial Voronoi bin with a single kinematic component parameterized by velocity $v$, dispersion $\sigma$, and Gauss--Hermite moments $h_3$ and $h_4$. Initial parameters are set to $v=0\,\kms$ (BCG rest frame), $\sigma=400\,\kms$, and $h_3=h_4=0$. We use the full set of SMILES templates \citep{Knowles2023}, including varying \aFe{}, to prioritize accurate recovery of the kinematics.

The resulting radial dispersion profile (Fig.\,\ref{fig:disp_h4}A) is approximately flat at ${\sim}350\,\kms$ in the BCG-dominated centre, increases beyond ${\sim}10$\,kpc, and then rises to ${\sim}750\,\kms$ at ${\sim}60$\,kpc. It continues to rise to ${\sim}1400\,\kms$ by ${\sim}100$\,kpc, but these points are not shown in Fig.\,\ref{fig:disp_h4}, since they fall below our surface brightness threshold for reliable kinematic measurements (Sect.\,\ref{sec:decomposition}). The $h_4$ (kurtosis) profile (Fig.\,\ref{fig:disp_h4}B) peaks at ${\sim}35$\,kpc, indicating a LOSVD with enhanced wings consistent with the superposition of a narrow (BCG-like) and a broad (ICL-like) component. These diagnostics motivate the subsequent approach of fitting multiple components to each spatial bin.

\subsection{Empirical BCG and ICL templates}\label{sec:BCGICLtemplates}

The S/N of the data is insufficient to robustly fit two stellar components with fully independent kinematics and stellar populations. We therefore make the simplifying assumption that the BCG and the ICL can each be represented by a single, fixed stellar population.

We construct empirical templates for the BCG and ICL stellar populations by extracting high-S/N spectra in regions we expect to be dominated by a single component. The BCG spectrum is extracted within a 0--5\,kpc circular aperture; the ICL spectrum is extracted from a 50--100\,kpc elliptical annulus (with $\epsilon=0.43$ and PA\,$=50$\,deg), after masking satellites and foreground/background objects. These regions are marked on Fig.\,\ref{fig:HST_MUSE_bins}.
We then fit these spectra using \ppxf{} \citep{Cappellari2023}, with SSP templates that allow variation in \aFe{} to mitigate template mismatch \citep{Knowles2023}.
To avoid contaminating our BCG template with projected ICL, when fitting the spectrum for the central region we include an additional component with fixed stellar population and LOSVD defined by the fit to the ICL region. The observed high-S/N spectra are shown in Fig.\,\ref{fig:spectra}, overlaid with the \ppxf{} models.

The best-fitting stellar population templates indicate that both the ICL and BCG components are old: the ICL and BCG have luminosity-weighted ages of ${\sim}9.3$ and ${\sim}9.6$\,Gyr, respectively. Their metallicities differ, with the BCG being metal-rich ($\langle\mathrm{[M/H]}\rangle=+0.26$) compared to the mildly supersolar ICL ($\langle\mathrm{[M/H]}\rangle=+0.048$). The inferred $\alpha$-enhancements are similar, given the uncertainties, with the ICL and BCG yielding $\langle[\alpha/\mathrm{Fe}]\rangle\simeq+0.03$ and $\simeq0.0$, respectively.  The best-fitting stellar population templates indicate that the BCG has a mass-to-light ratio of $8.4\,\mathrm{M}_{\odot}/\mathrm{L}_{\odot}$, whereas the ICL has a lower mass-to-light ratio of $5\,\mathrm{M}_{\odot}/\mathrm{L}_{\odot}$.
The contrasting spectral templates, resulting from these stellar-population differences, help us to decompose the BCG and ICL in regions where they have similar surface brightnesses.

\subsection{Multi-component decomposition}\label{sec:decomposition}

We now model the spectrum of each spatial Voronoi bin as the sum of two kinematic components with independent velocities, dispersions, and relative weights. One template is always fixed to the BCG stellar population and the other is fixed to the ICL stellar population template. The baseline model contains BCG and ICL components with initial line-of-sight $v=0\,\kms$, and with initial dispersions of $\sigma_{\rm BCG}=357$ and $\sigma_{\rm ICL}=791\,\kms$ (as measured from the high-S/N BCG and ICL spectra used to construct the empirical stellar population templates). Bounds of $\pm3000\,\kms$ in velocity and $0$--$2000\,\kms$ in dispersion are imposed. We find that the decomposition becomes unreliable for spatial bins with a total surface brightness fainter than $25$\,mag\,arcsec$^{-2}$, so we only make use of bins brighter than this limit in our kinematic analysis. These bins are marked by a yellow contour in Fig.\,\ref{fig:models}C and F. For defining this threshold, the total surface brightness in each bin is estimated by convolving our best-fitting single-component model with the \textit{HST} F606W transmission curve.

\begin{figure}
    \centering
    \includegraphics[width=\linewidth]{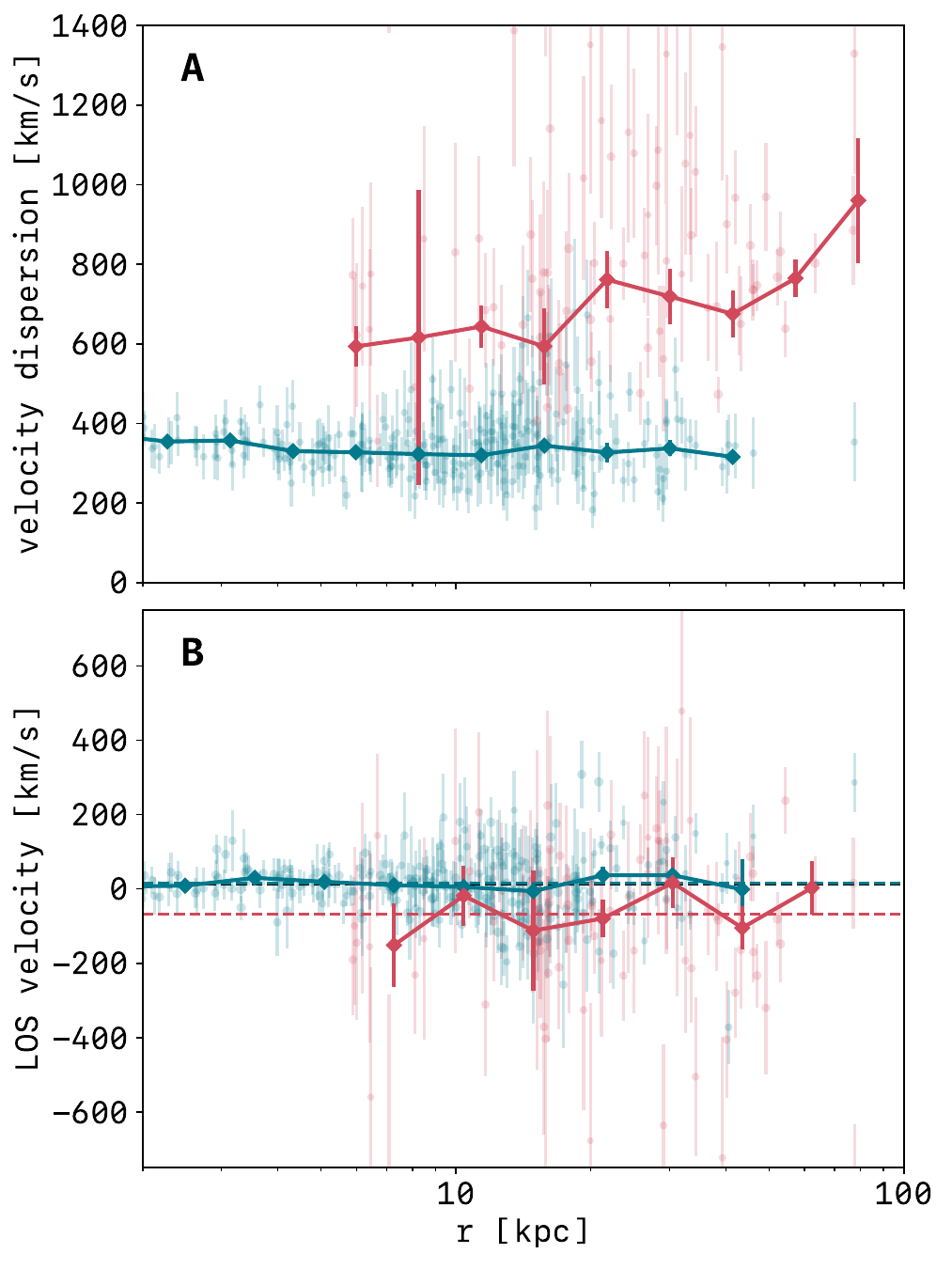}
    \caption{The kinematics of the BCG and ICL components versus clustercentric radius.
    The velocity dispersion (\textbf{A}) and line-of-sight velocity (\textbf{B}) of the BCG (blue) and ICL (red), from two-component fits to our spatially binned MUSE data.
    Small points show the \ppxf{} measurement, with its uncertainty, for the individual spatial bins. To ensure our kinematic decomposition is robust, we only consider bins where the total (BCG+ICL) surface brightness is brighter than 25\,mag\,arcsec$^{-2}$ and the uncertainty on the velocity dispersion is less than one third of its measured value. The large diamonds show the error-weighted mean value in logarithmic bins of radius, with error bars indicating the standard error within each radial bin.
    The dashed lines in panel \textbf{B} indicate the error-weighted mean velocities of each component.
    }
    \label{fig:disp_vel}
\end{figure}

The kinematic properties of the two components are shown in Fig.\,\ref{fig:disp_vel}.
To explore how different starting values can affect our results, we tested several different initial velocity dispersions for both components and find similar results.
In Fig.\,\ref{fig:disp_vel}A we see that the kinematic component corresponding to the BCG stellar population has an almost flat velocity-dispersion profile at all radii, with a mean value of $335\pm4\,\kms$.
The ICL component likewise displays an approximately flat velocity-dispersion profile, with a mean of $690\pm30\,\kms$. The near constancy of both dispersions over an order of magnitude in radius strengthens our picture of a dynamically hot ICL component that penetrates into the inner regions of the cluster while continuing to trace the cluster's large-scale mass distribution. In addition to the differing velocity dispersions, the two components are also offset in line-of-sight velocity (see Fig.\,\ref{fig:disp_vel}B), with the BCG moving at $86\pm18\,\kms$ relative to the ICL.

%AS1063 hosts a known low-velocity galaxy substructure, with a secondary peak around $-2000\,\kms$ relative to the BCG \citep{Mercurio2021}. In early stages of our analysis we found that regions of diffuse light to the south-west of the BCG exhibited similarly blueshifted velocities. This motivated including an associated intragroup light (IGL) component in our model. However, after further development of our background and masking algorithms (Sect.\,\ref{sec:background} and \ref{sec:masking}), an additional IGL kinematic component is no longer required: a two-component BCG+ICL model is sufficient.

We next explore the spatial distribution of the ICL and BCG components. To improve the robustness of the ICL fluxes in the cluster centre, where the BCG dominates, we perform a second set of \ppxf{} fits in which we fix both the stellar population templates (BCG and ICL, as above) and also the kinematics of each component to their average values derived above. The BCG kinematics are fixed to a line-of-sight velocity of $v=16\,\kms$ and $\sigma_{\rm BCG}=335\,\kms$, and the ICL is fixed to $v=-69\,\kms$ with $\sigma_{\rm ICL}=691\,\kms$. In these fits, only the relative weights of the two components are allowed to vary.
We reconstruct a flux-calibrated spectrum for each component in every spatial bin, and convolve these spectra with the \textit{HST} F606W response curve to obtain the decomposed BCG and ICL surface brightness in each bin.

We again limit our use of a two-component model to bins with a F606W surface brightness of at least $25$\,mag\,arcsec$^{-2}$. Although this second decomposition is more robust, at fainter surface brightness levels we find the contribution of the BCG is negligible (see Fig.\,\ref{fig:3prof}). In fainter bins, we therefore set the BCG component to zero and estimate the ICL surface brightness from a single-component fit using the ICL template (with fixed kinematics). The resulting surface brightness maps (panels C and F of Fig.\,\ref{fig:models}) reveal that the BCG and ICL components have contrasting spatial distributions.

\section{Surface brightness and ICL fraction}\label{sec:SBmaps}

\subsection{One-dimensional surface brightness profiles}\label{sec:1dsurfacebrightness}

To measure decomposed one-dimensional surface brightness profiles of the BCG and ICL, we first convert the radii of each bin to the equivalent semimajor axis. This gives a more meaningful comparison than circular radii, which intersect bins of significantly different surface brightnesses.

The ellipticity of the cluster's stellar component varies with semimajor axis, therefore we used \texttt{AutoProf} \citep{Stone2021} to measure the variation of the ellipticity and position angle in the \textit{HST} F606W image. This was used to create a map assigning each spatial bin an equivalent semimajor axis, corresponding to the semimajor axis of an isophote which passes through that bin. The same mapping was used for both the BCG and ICL, allowing a direct comparison of the one-dimensional surface brightness profiles.

\begin{figure}
    \centering
    \includegraphics[width=\linewidth]{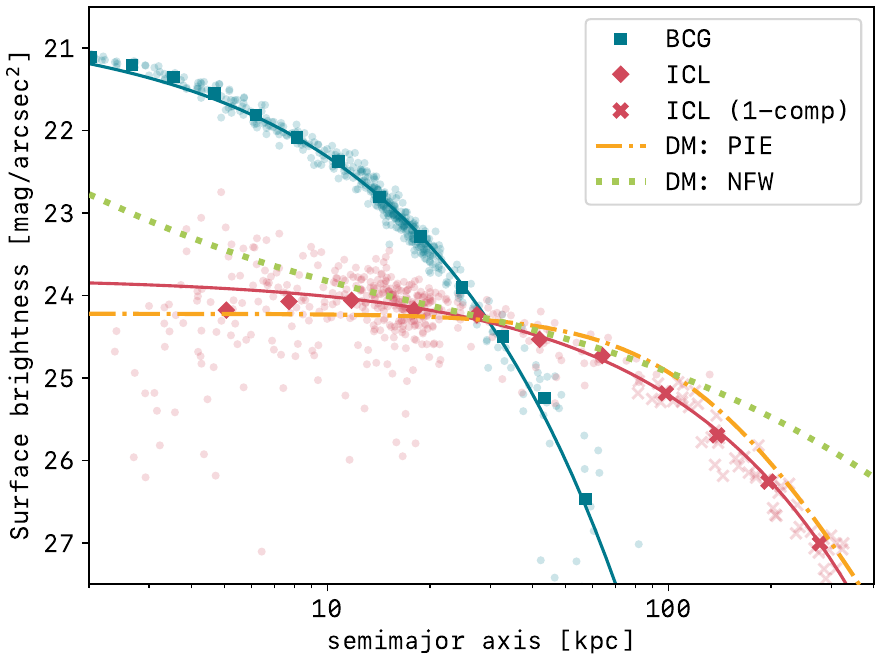}
    \caption{One-dimensional surface-brightness profiles of the BCG and ICL.
    The surface brightness of each component in the MUSE spectra convolved with the \textit{HST} F606W transmission curve, plotted against the semimajor axis of the isophote that passes through the point. Small blue (BCG) and red (ICL) points show individual spatial bins; large symbols show the mean values in logarithmic bins of semimajor axis. Solid curves show a one-dimensional S\'ersic fit to the large points for each component. Crosses indicate bins where we assign all the flux to the ICL. Over-plotted are two dark matter surface mass profiles normalised to the ICL at 30\,kpc: a pseudo-isothermal ellipse (PIE) profile with a core radius of 100\,kpc, and an NFW profile with an inner slope of $\gamma=1$. The ICL is a close match to the cored, pseudo-isothermal model.}
    \label{fig:3prof}
\end{figure}

The surface brightness of each component within each spatial bin in plotted as a function of semimajor axis in Fig.\,\ref{fig:3prof}. We recall that, apart from defining the BCG and ICL templates from the central and outer regions of the MUSE field, we have not imposed any prior spatial distribution on either component. We find that the BCG dominates the light within 25\,kpc, with the light beyond 50\,kpc being entirely ICL. Nevertheless, the surface brightness of the ICL component increases monotonically toward the centre of the cluster. Even though the BCG component is 2.5\,mag brighter at a radius of ${\sim}5$\,kpc, we still significantly detect the presence of an ICL component. The transition region between the BCG and ICL components lies between 30--40\,kpc, beyond which the ICL dominates the light profile.

To interpolate the decomposed surface brightness profiles, we fit S\'ersic profiles to the mean surface brightness as a function of semimajor axis. The resulting profiles are shown by the solid lines in Fig.\,\ref{fig:3prof}. The spectroscopically decomposed BCG and ICL components closely resemble distinct S\'ersic profiles.

\subsection{Two-dimensional surface brightness maps}\label{sec:2Dfitting}

\begin{figure*}
    \centering
    \includegraphics[width=\hsize]{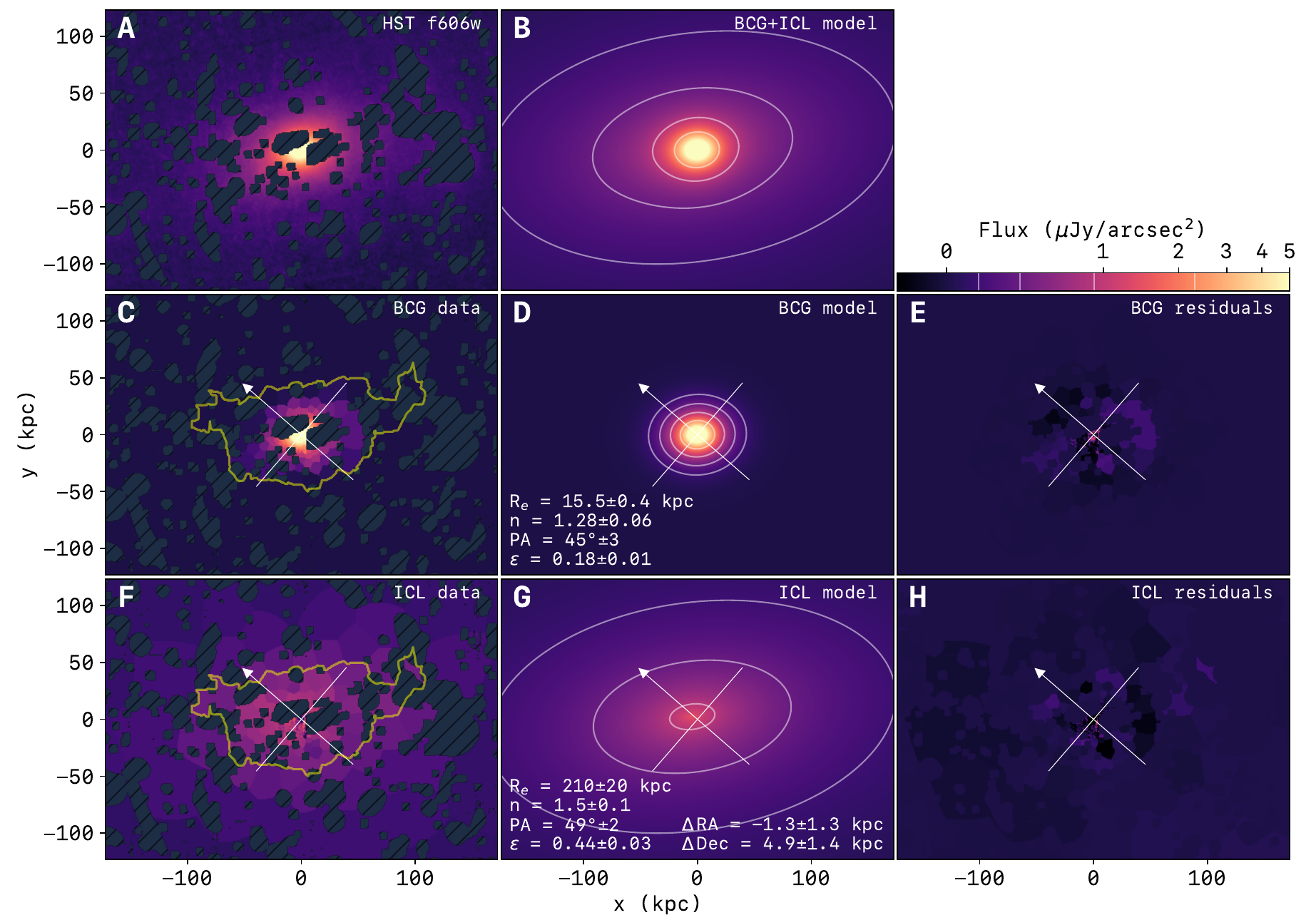}
    \caption{The distribution of the BCG and ICL in each spatial bin of the MUSE datacube.
\textbf{A}:~\textit{HST} F606W image smoothed with a Gaussian of width $\sigma = 0.55$\,arcsec to mimic the seeing of the MUSE observations; all sources other than the BCG and ICL are masked (grey regions).
\textbf{B}: The total BCG+ICL light obtained by summing the individual models shown in panels D and G.
\textbf{C}:~Image constructed from the MUSE data showing the light attributed to the BCG in each spatial bin.
The yellow contour encloses bins brighter than 25\,mag\,arcsec$^{-2}$, for which we can separate the BCG and ICL components; beyond this we assign all the MUSE flux to the ICL.
The arrow points North, and East is anticlockwise.
\textbf{D}: S\'ersic model for the BCG light in panel C. Best-fit values and $1\sigma$ uncertainties are given for the half-light radius ($R_{\mathrm{e}}$), S\'ersic index ($n$), position angle (PA; degrees East of North), and ellipticity ($\epsilon = 1 - b/a$).
\textbf{E}:~Residuals between the BCG model (panel D spatially binned) and the data in panel C.
\textbf{F}:~As panel C, but showing the light attributed to the ICL in each spatial bin.
\textbf{G}: S\'ersic model for the ICL light in panel F. RA and Dec offsets are given relative to the BCG, defining West as positive.
\textbf{H}:~Residuals between the ICL model (panel G spatially binned) and the data in panel F.}
    \label{fig:models}
\end{figure*}

In the \textit{HST} images, diffuse ICL extends into the same region that we use to define our background spectrum. We therefore accept that our MUSE background correction will over-subtract some light from each spectrum. When performing the surface brightness fitting of the ICL, we counter this over-subtraction by adding a constant of $0.03\,\mu$Jy\,arcsec$^{-2}$ into the modelled two-dimensional ICL distribution (Fig.\,\ref{fig:models}F) before measuring its S\'ersic parameters. This corresponds to the surface brightness of the ICL measured in the smoothed \textit{HST} F606W image in the same region over which we estimate our MUSE background level. We do not add any constant to the BCG image (Fig.\,\ref{fig:models}C), since Fig.\,\ref{fig:3prof} shows that the BCG surface brightness at such large radii is negligible.

To quantify the spatial structure of the BCG and ICL, we fit independent two-dimensional elliptical S\'ersic profiles to their surface brightness distributions, using a modified version of the \textsc{AstroPhot} package \citep{Stone2023}. For each component, the free parameters are the centroid ($\Delta$RA, $\Delta$Dec), ellipticity ($\epsilon$), position angle (PA, measured in degrees East of North), half-light radius ($R_{\mathrm{e}}$), and S\'ersic index ($n$).
We estimate the flux uncertainty in each spatial bin from the F606W-convolved MUSE data and scale it by the fraction of flux assigned to the component (BCG or ICL) in that bin. Handling the Voronoi-binned data required some small modifications to \textsc{AstroPhot}. After each trial model is sampled onto a regular pixel grid, it is binned to the same Voronoi tessellation as the data. The residuals are then determined for each spatial bin, and the optimization proceeds in the usual manner. Reliable uncertainties for the component parameters are estimated by taking the standard deviation over independent fits to many bootstrap resamplings of the spatial bins.

The resulting models for the BCG and ICL are shown in panels D and G of Fig.\,\ref{fig:models}, and their sum (panel B) reproduces the masked \textit{HST} image (panel A) remarkably well. Although the ICL flux in each spatial bin is determined independently, its two-dimensional distribution is nonetheless well described, over an order of magnitude in radius, by a single elliptical S\'ersic profile. This inspires confidence that these fits provide tight constraints on the radial profile, ellipticity and projected centre of the ICL.

\begin{figure}
    \centering
    \includegraphics[clip, trim=1.5cm 0.2cm 1.8cm 0.05cm, width=\linewidth]{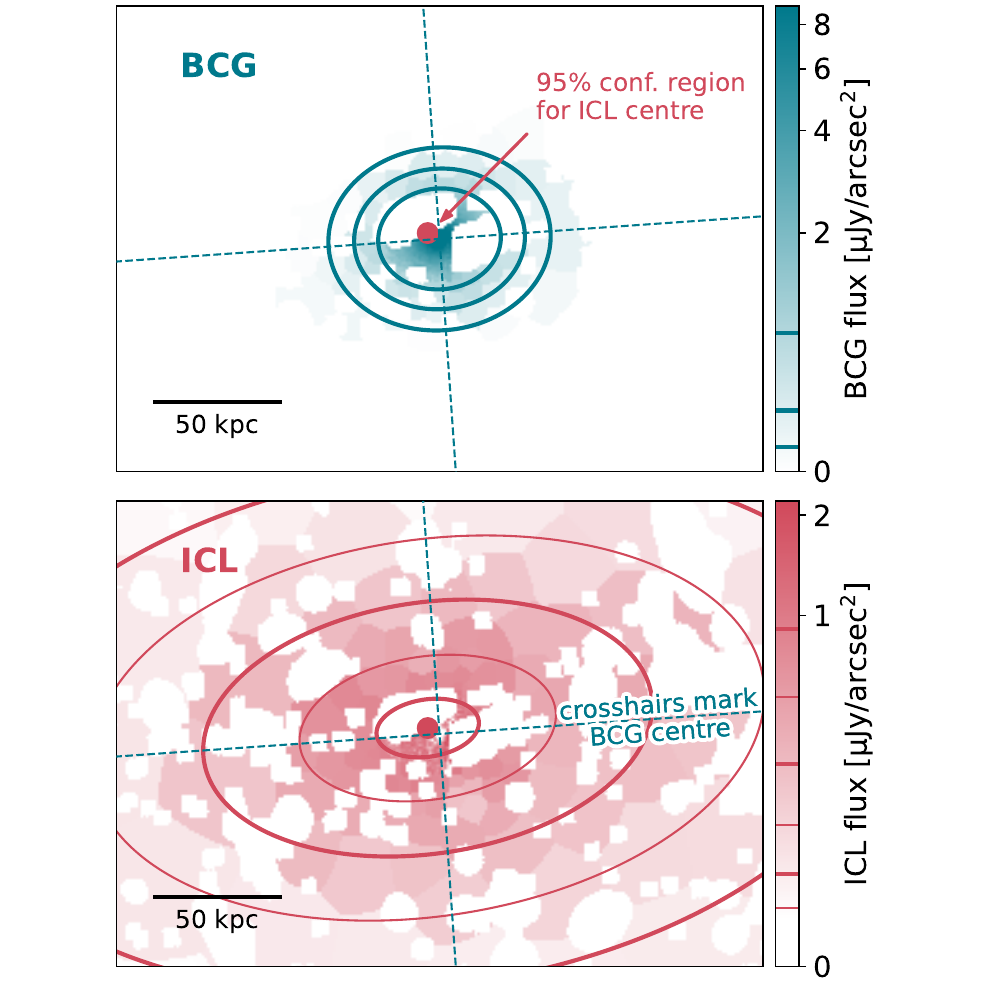}
    \caption{Centres of the spectroscopically separated BCG and ICL components in AS1063. The panels show the masked and binned flux attributed to the BCG (left) and ICL (right), with elliptical isophotes from the S\'ersic fits overlaid. The fitted centres are marked; they are offset by $5.2\pm1.5$\,kpc, with uncertainties estimated by bootstrap resampling the spatial bins.}
    \label{fig:centres}
\end{figure}

The BCG has a half-light radius of $15.5\pm0.4$\,kpc and an ellipticity of $0.18\pm0.01$, typical of massive cluster elliptical galaxies \citep{Shen2003, Holden2009}. By contrast, the ICL is more extended and elongated, with a half-light radius of $210\pm20$\,kpc and ellipticity $0.44\pm0.03$, and its centroid is offset from the BCG by $5.2\pm1.5$\,kpc (as shown more clearly in Fig.\,\ref{fig:centres}).

The fitted BCG component has a S\'ersic index $n \simeq 1.3$, which may seem low for a massive elliptical. However, the traditional $n\sim4$ profiles of such galaxies arise, at least in part, from substantial flux at large radii that our spectral decomposition assigns to the ICL. Some flattening of the central light profile by the MUSE PSF may also contribute.

\subsection{ICL fraction}\label{sec:ICLfraction}

The ICL fraction provides a useful metric for comparisons within and between studies. We first consider the fraction of the flux (or stellar mass) in the ICL divided by the combined BCG+ICL, i.e. without including satellite galaxies in the denominator. We calculate this quantity using the S\'ersic models derived from the two-dimensional surface brightness maps in Fig.\,\ref{fig:models}, which correspond to the F606W filter. Within a circular aperture of 100\,kpc the ICL/(BCG+ICL) light fraction is 67\,per cent. Converting light to stellar mass using the mass-to-light ratios from our template fitting (BCG $8.4\,\mathrm{M}_{\odot}/\mathrm{L}_{\odot}$; ICL $5\,\mathrm{M}_{\odot}/\mathrm{L}_{\odot}$) reduces this fraction to 55\,per cent.

Extending the aperture to the maximum radius of our observations, 300\,kpc, we estimate the BCG emits $1.8\times10^{11}$\,L$_{\odot}$ from $1.5\times10^{12}$\,M$_{\odot}$ of stars, whereas the ICL emits $9.1\times10^{11}$\,L$_{\odot}$ from $4.5\times10^{12}$\,M$_{\odot}$ of stars. This gives an ICL/(BCG+ICL) light fraction of 83\,per cent and mass fraction of 75\,per cent.

To estimate the ICL fraction in terms of the total stellar mass in the cluster, we can use the galaxy stellar mass determined by \citet{Morishita2017}. They measure a total stellar mass of $9.8\times10^{11}$\,M$_{\odot}$ within galaxies (other than the BCG), assuming the same initial mass function as in this work. Thus, the ICL constitutes 64\,per cent of the total stellar mass within 300\,kpc of the cluster centre.

\section{Discussion}\label{sec:discussion}

\subsection{Properties of the ICL}

The luminosity-weighted ages of the BCG and ICL templates (${\sim}9.6$ and ${\sim}9.3$\,Gyr) indicate that both populations formed at early times. The factor of ${\sim}1.6$ difference in metallicity -- supersolar in the BCG, near solar in the ICL -- is consistent with previous studies of ICL metallicity \citep{Montes2014, Montes2018, Edwards2020, Barbosa2021} and suggests that BCG stars predominantly formed in the centres of massive galaxies, while ICL stars originate mainly in low-mass systems or the outskirts of massive galaxies \citep{Brown2026}. The similar, near-solar \aFe{} values fall within the range observed for massive BCGs and their halos \citep[e.g.][]{Gu2018}, and may reflect similar enrichment timescales for the dominant progenitors of both components.

From the high-S/N BCG and ICL spectra we used to construct the empirical stellar population templates, we measure velocity dispersions of $\sigma_{\rm BCG}=357$ and $\sigma_{\rm ICL}=791\,\kms$. Furthermore, our two-component decomposition yields mean values of $\sigma_{\rm BCG}=335\pm4\,\kms$ and $\sigma_{\rm ICL}=690\pm30\,\kms$, both approximately flat with over the radii where we have measurements, $5$--$100$\,kpc. The ICL velocity dispersion is therefore consistently more than twice that of the BCG, supporting our assumption of two distinct physical components pervading the cluster core.

Our measured ICL velocity dispersion is only around half that of the cluster galaxies ($1380^{+26}_{-32}\,\kms$) measured by \citet{Mercurio2021}. However, in simulations, \citet{Butler2025} find that the orbital energy of the ICL is around 30\,per cent lower than that of the satellite galaxies, corresponding to a substantially lower velocity dispersion. \citet{Marini2021} scale their simulations to match the halo properties of AS1063 and estimate a stellar velocity dispersion of ${\sim}300\,\kms$ in the BCG and ${\sim}700\,\kms$ at a radius of $100$\,kpc, which are very similar to our measured values. The transition region between our BCG and ICL components lies around $35$\,kpc, within the range of $60\pm40$\,kpc predicted in simulations by \citet{Contini2022}. Our kinematic separation is thus in excellent agreement with expectations.

Our kinematically constrained ICL profile differs markedly from the photometric decomposition of this system by \citet{Morishita2017}, who attribute considerably more light to the BCG at all radii. Within 100\,kpc we find that 55\,per cent of the BCG+ICL stellar mass belongs to the ICL, significantly larger than the 29\,per cent inferred from the photometric analysis. Within 300\,kpc, the ICL constitutes 64\,per cent of the total stellar mass (including other galaxies), compared to the 23\,per cent obtained from their purely photometric decomposition. Our method shows, for the first time in observations -- but as predicted by simulations \citep{Brough2024} -- that photometric measures of the ICL underestimate the amount of stars that orbit in the cluster halo, rather than being bound to any particular galaxy.

\subsection{The ICL as a tracer of dark matter}

To evaluate the ICL as a tracer of the dark matter distribution, we compare its shape, centroid and radial profile with those of the dark matter distribution inferred from strong gravitational lensing \citep{Caminha2016}. Although strong lensing has its own limitations, it is currently the best available tool for estimating the dark matter distribution in individual clusters.

The position angle of the ICL is consistent, within the uncertainties, with that of the lensing-derived dark matter halo \citep[table~5 of][]{Caminha2016}. The ICL is significantly rounder (ellipticity $\epsilon=0.44$) than the dark matter halo ($\epsilon=0.6$) in the lensing model, in contrast with findings by \citet{Gassis2025} that the ICL halos of relaxed simulated clusters are typically more elongated than their dark matter halos, albeit with a large scatter. However, the dark matter model of AS1063 assumes a single ellipticity for the entire cluster, whereas simulations indicate that dark matter cores are typically rounder than halo outskirts \citep{Fernadez2025}, which may explain this discrepancy.

The detected centroid offset of the ICL from the BCG (Fig.\,\ref{fig:centres}), of $1.3\pm1.3$\,kpc East and $4.9\pm1.4$\,kpc North, is similar to the offset of the lensing mass centroid, $2.5\pm1.1$\,kpc East and $2.6\pm0.7$\,kpc North.
In summary, the ICL shape and centroid closely match the strong-lensing estimate of the underlying dark matter halo, with the rounder ICL ellipticity plausibly closer to the true shape of the inner halo. This supports the view that isolating the ICL from the BCG allows us to probe the two-dimensional shape of the cluster's central dark matter potential.

Turning to the radial profile, our constraints on the ICL light profile and kinematics together point to a total mass distribution that is well described by a cored halo. Fig.\,\ref{fig:3prof} illustrates that the ICL profile within the inner ${\sim}300$\,kpc closely matches the shape of a cored dark matter profile. Furthermore, the ICL velocity dispersion is high (${\sim}690\,\kms$) and approximately flat from 5 to 100\,kpc, indicating that the inner potential varies only slowly with radius. Both signatures support a cored dark matter halo in this cluster. This is consistent with the strong-lensing analysis of \citet{Caminha2016}, which favours a pseudo-isothermal halo with a ${\sim}100$\,kpc core over a cuspy NFW \citep{NFW} profile.

Further evidence that the centre of this cluster has a relatively flat density profile comes from the observed offset in phase space ($5.2\pm1.5$\,kpc and $86\pm18\,\kms$) between the ICL and the BCG. This indicates that, while the BCG is bound to the cluster core, it nonetheless oscillates within a relatively broad minimum of the dark matter potential. While such oscillations are expected if the cluster has experienced a recent merger, they would be rapidly damped by the cuspy core predicted in standard CDM, but can persist for longer in a cored dark matter halo, such as those predicted by simulations featuring SIDM (e.g., \citealt{Harvey2025}). Spectroscopic isolation of the ICL provides both spatial and line-of-sight velocity offsets between the BCG and the large-scale halo tracer -- i.e.\ a phase-space offset -- whereas most other methods measure only the projected spatial offset. An anti-correlation between spatial and velocity offsets is a hallmark of harmonic oscillation, and hence of the cored halo \citep{Harvey2025}.

The BCG--halo offset in any given cluster can be biased by its particular dynamical state and environment -- for example, a recent merger can excite a temporary wobble even in a cuspy halo. In AS1063, there is evidence for recent merging activity, from the cluster's elongated X-ray emission \citep{Gomez2012} and the bimodality of the galaxy velocity distribution \citep{Mercurio2021}. 
However, while a single detection does not by itself challenge CDM, a statistical sample of BCG--ICL phase-space offsets, compared to expectations from simulations featuring different dark matter models, could potentially provide constraints on the nature of dark matter \citep{Harvey2019}.

\section{Conclusions}\label{sec:conclusions}

We have demonstrated that integral-field spectroscopy, combined with empirical stellar-population templates, can isolate the BCG and ICL in AS1063 and measure their distinct kinematics, spatial distributions, and stellar populations -- opening a route both to exploring assembly history of cluster cores and revealing the shape of the inner dark matter halo. Our main conclusions are as follows.

\begin{enumerate}
\item A two-component spectral decomposition yields nearly flat velocity-dispersion profiles for the BCG ($335\pm4\,\kms$) and ICL ($690\pm30\,\kms$), with a line-of-sight velocity offset of $86\pm18\,\kms$.
\item The BCG and ICL are both old (${\sim}9$\,Gyr), but the BCG is more metal-rich, consistent with BCG stars forming in the centres of massive galaxies and ICL stars originating in lower-mass systems or galactic outskirts.
\item The spectroscopically isolated components are well described by distinct S\'ersic profiles. Within 100\,kpc, the ICL comprises 55\,per cent of the BCG+ICL stellar mass -- substantially more than photometric decompositions of the same system imply -- in line with simulation predictions that photometry underestimates the kinematically defined ICL.
\item Two-dimensional S\'ersic fits give an ICL half-light radius of $210\pm20$\,kpc, ellipticity $0.44\pm0.03$, and a centroid offset of $5.2\pm1.5$\,kpc from the BCG. The ICL shape and centroid agree with strong-lensing estimates of the dark matter halo.
\item The ICL surface-brightness profile, flat high velocity dispersion, and BCG--ICL phase-space offset are consistent with a cored dark matter halo in AS1063, matching the PIE model preferred by \citet{Caminha2016}. A single cluster cannot discriminate among dark matter models. However, applied to a statistical sample, the joint distribution of spatial and kinematic offsets -- together with ICL surface-brightness profiles -- offers a complementary dynamical probe of cluster halo cores, with distinct advantages over lensing and X-rays alone, and a new avenue for testing the nature of dark matter.
\end{enumerate}

\section*{Acknowledgements}

NAH gratefully acknowledges support from the Leverhulme Trust through a Research Leadership Award. CB, NAH, and SB acknowledge support from the UK Science and Technology Facilities Council (STFC) under grant ST/X000982/1.

%%%%%%%%%%%%%%%%%%%%%%%%%%%%%%%%%%%%%%%%%%%%%%%%%%
\section*{Data Availability}

The data underlying this article were accessed from the ESO public archive and the HST Frontier Fields archive. The derived data generated in this research will be shared on reasonable request to the corresponding author.

%%%%%%%%%%%%%%%%%%%% REFERENCES %%%%%%%%%%%%%%%%%%

\bibliographystyle{mnras}
\bibliography{ref}

%%%%%%%%%%%%%%%%%%%%%%%%%%%%%%%%%%%%%%%%%%%%%%%%%%

% Don't change these lines
\bsp	% typesetting comment
\label{lastpage}
\end{document}